\documentclass[prb,aps,twocolumn,showpacs]{revtex4}
\usepackage{graphicx}
\usepackage{bm}
\usepackage{amsmath}
\usepackage{color}

\begin{document}
\title{Electric field control of spin-orbit splittings in GaAs/AlGaAs coupled quantum wells}
\author{A.~V.~Larionov$^1$\email{larionov@issp.ac.ru}}
\author{L.~E.~Golub$^2$}
\affiliation{$^1$Institute of Solid State
Physics, Russian Academy of Sciences, 142432 Chernogolovka, Russia}
\affiliation{$^2$A.~F.~Ioffe Physico-Technical Institute, Russian
Academy of Sciences, 194021 St. Petersburg, Russia}

%\date{\today}
\begin{abstract}
Electron spin dynamics is investigated in n-i-n GaAs/AlGaAs coupled quantum wells. 
The electron spin dephasing time is measured as a function of an external electrical bias under resonant excitation of the 1sHH intrawell exciton using a time-resolved Kerr rotation technique. 
It is found a strong electron spin dephasing time anisotropy caused by an interference of the structure inversion asymmetry and the bulk inversion asymmetry. 
This anisotropy is shown to be controlled by an electrical bias. 
A theoretical analysis of electron spin dephasing time anisotropy is developed. 
The ratio of Rashba and Dresselhaus spin splittings is studied as a function of applied bias.
\end{abstract}
\pacs{73.21.Fg, 73.63.Hs, 72.25.Rb, 76.60.Jx}

\maketitle

{\bf Introduction.} Critical point of spintronic
investigations is a control of spin degrees of freedom by
electrical means. There are proposals to create electronic devices
working due to an electric field effect on the orientation of
electron spins. In order to study such phenomena one needs to use
a coupling between orbital and spin degrees of freedom. This is
possible due to spin-orbit interaction, an universal relativistic
effect which, however, can be altered in semiconductors and
low-dimensional semiconductor systems by special structure design
and/or external parameters. Especially interesting for both
fundamental physics and applications is the spin-orbit interaction
caused by lack of inversion center in the system. The important
example is a class of effects caused by Structure Inversion
Asymmetry (SIA) which is present in two-dimensional
semiconductor structures.~\cite{spintronics}
%The first spintronic proposal, the spin field effect transistor, is
%based on the manipulation by SIA via external electric field.

There is a number of works where the SIA degree has been changed
in two-dimensional structures by an external gate.~\cite{Nitta,Karimov03}
However in most cases the effect of the gate voltage is a change
of the electron gas concentration, which produces an additional
internal electric field affecting SIA. 
Among the quasi-two-dimensional objects based on semiconductor
heterostructures, Coupled Quantum Wells (CQWs) are of special
interest. Electrical bias in such structures does not produce
extra carriers but has dramatic effects on SIA. This allows for direct
manipulation by spin-orbit interaction even in undoped structures.
In addition, CQWs are very suitable objects for spin dynamics
study because, due to spatial separation of photoexcited electrons
and holes in neighboring quantum wells, the radiative lifetimes
are long enough so that the spin lifetime is determined by spin
relaxation processes.

SIA manifests itself as a source of spin relaxation of free
electrons in two-dimensional semiconductors. It generates an
effective magnetic field rotating electron spins which is
the basis of the D'yakonov-Perel' spin relaxation mechanism, for review see Ref.~\onlinecite{Averkiev02}.
Accordingly, the spin relaxation times measurements give the
necessary information about the degree of SIA.
In addition to SIA, there is another source for lacking inversion
symmetry in  semiconductors. This is Bulk Inversion
Asymmetry (BIA) present in structures based on
noncentrosymmetric materials and also leading to the
D'yakonov-Perel' spin relaxation. If both SIA and BIA  are
present, new interesting effects appear in spin dynamics. In
particular, the spin relaxation anisotropy has been predicted for
a spin oriented in the plane of a (001) grown
structure.~\cite{Averkiev99} It has been shown
that the anisotropy should change dramatically if the SIA strength is
tuned to be comparable to BIA. However, in
Refs.~\onlinecite{Averkiev06,Stich} the anisotropy has been
observed for fixed sets of parameters, and in
Ref.~\onlinecite{APL_china} it has been changed varying the
electron concentration by other means.
Effect of electrical bias on spin relaxation has been observed due
to SIA variation but in specifically oriented (110)
structures~\cite{Karimov03} where SIA and BIA do not interfere and
the in-plane anisotropy is absent, or when spin relaxation has not
been caused by the spin-orbit interaction at
all.~\cite{Gerlovin07}

In the present work we use biased CQW structures for smooth change of 
spin relaxation anisotropy  by electrical means. We demonstrate 
that spin-orbit interaction can be controlled by electric field in n-i-n GaAs/AlGaAs CQWs.

Time resolved Kerr rotation spectroscopy is an
effective technique for creation and study of spin coherence in
semiconductors.~\cite{spintronics} It allows one to prepare a coherent
superposition of electron and (or) hole basis states. 
When external
magnetic field is applied perpendicular to the direction of the
circularly polarized light propagation, 
the spin vector  precesses in the plane normal to the applied field. From a
quantum-mechanical point of view, this Larmor precession
corresponds to quantum beats (QBs) between spin-splitted states of the Zeeman
doublet.
A projection of the
spin vector on its excitation direction oscillates 
and, simultaneously, decays due to decoherence. The
decoherence is caused by spin relaxation  of electrons and (or) holes which allows one to study various spin relaxation mechanisms by time resolved Kerr rotation measurements.

{\bf Experiment.}
The CQW system studied here consists of two $120$~\AA  \mbox{} wide GaAs quantum wells
with a narrow (4 monolayer) AlAs barrier between them. The quantum wells are separated from the Si-doped
($10^{18}$cm$^{-3}$) GaAs layers by 0.15~$\mu$m thick
Al$_{0.33}$Ga$_{0.67}$As barriers. The upper part of the structure
is covered with a $100$~\AA \mbox{} GaAs layer. Mesa structures
were fabricated on the as-grown structure by a lithographic
technique. Metal contacts of Au + Ge + Pt alloy were deposited as
a frame on the upper part of the mesa and also the doped buffer
layer.

The setup for Kerr rotation measurements consists
of a high precision mechanical delay line (OWIS, LIMES 170) and a
photo-elastic modulator (PEM, I/FS50). The detection of nonlinear signals is provided by a Si p-i-n photodiodes balanced
detector (Nirvana-2007) combined with a lock-in detector. As a
source of pulsed photoexcitation we have used a femtosecond
Ti-Sapphire laser (Tsunami) with pulse shaping up to about 2~ps.
\begin{figure}[t]
\begin{center}
\includegraphics[width=0.9\linewidth]{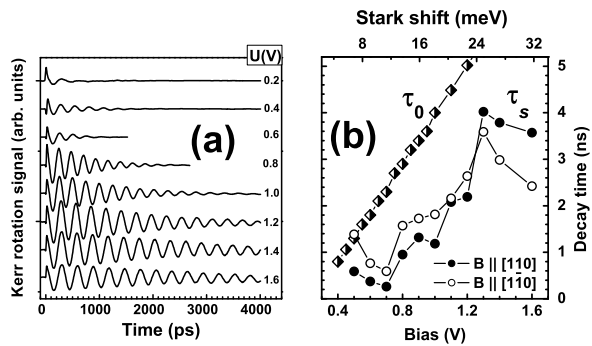}
\end{center}
\caption{(a) Kerr rotation signal at different applied biases measured at $B=1$~T (${\bm B} \parallel [1 1 0]$), temperature $\approx 2~$K and photoexcitation power density $\approx 2\times10^{3}$~W/cm$^{2}$.
(b) Electric field dependence of the photoluminescence decay time and the electron spin dephasing time at
two magnetic field orientations.
The Stark shift is given for the indirect exciton photoluminescence line.
}
\end{figure}
The CQW sample under study was mounted in magnet
cryostat with a split-coil superconducting solenoid generating
magnetic fields up to $B=6$~T at 2~K.

A set of time-resolved spin QBs detected in a magnetic field
of 1~T in the Voigt configuration is given in Fig.~1(a) for different
biases. The energy of pump and probe Ti-Sapphire pulsed laser beams
was the same and has been set to the maximum of the
photoluminescence line contour corresponding to radiative
annihilation of the 1sHH exciton. 
The experimental data can be fitted by an exponentially
damped oscillation containing the beating frequency $\Omega$ and
a single decay time:
\begin{equation}
\label{I}
I(t) = I_{0} \exp{(-t/T)} \cos{\Omega t}.
\end{equation}

The observed periodic
oscillations are due to precession of coherently excited
electron spins around the external magnetic field. The period of
the oscillations is proportional to the conduction electron Zeeman splitting. 
We have checked that the beating frequency $\Omega$ is linear in $B$ down to $B=0$. 
The in-plane $g$-factor is  $|g_{e}| \approx 0.25$ showing some variations with the bias as well as a weak in-plane anisotropy. This data will be published elsewhere. Note that the heavy-hole in-plane $g$-factor is an order of magnitude smaller.~\cite{Marie98} 

As a result of fitting by Eq.~\eqref{I} the dependence of the decay time $T$ on the applied bias is obtained. We have verified that $T$ is independent of the magnetic field strength. 
The decay rate is a sum of radiative recombination and spin dephasing rates:
$$1/T = 1/\tau_0 + 1/\tau_s.$$
We extracted the interwell radiative electron-hole annihilation time $\tau_0$ from the photoluminescence measurements (the intrawell one is rather short being of order of tens picoseconds). %This allowed us to get the spin dephasing time $\tau_s$. 
The dependences of both $\tau_s$ and $\tau_0$ on the applied bias are shown in Fig.~1(b).
One can see that $\tau_s$ is shorter than $\tau_0$ for $U>0.4$~V.
\begin{figure}[t]
\begin{center}
\includegraphics[width=0.8\linewidth]{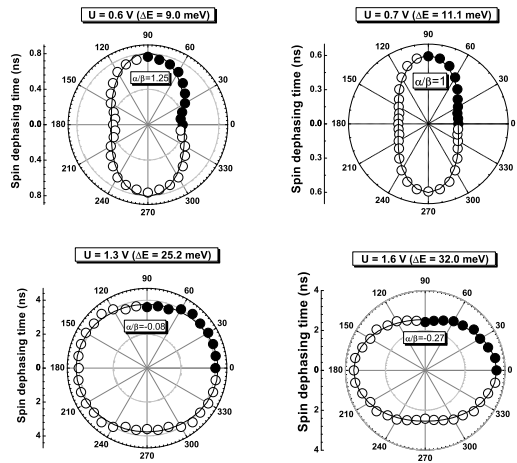}
\end{center}
\caption{Spin dephasing time measured as a function of the angle
between ${\bm B}$ and the axis $[1 1 0]$ at four
applied biases. Experimental data are shown by points, theory is presented by solid lines. 
The values $\Delta E$ represent the Stark shifts for the given biases.
}
\end{figure}

Figure~2 demonstrates the polar plot of the spin dephasing time vs the magnetic field orientation in the structure plane for four applied biases.
For such investigation it is necessary to rotate the sample relative to
the direction of applied magnetic field with a high accuracy.
Experimentally we have measured the data for angles from 0 to 90
degrees only (black points). Open points have been obtained by
extrapolation to the next three quadrants. One can see that
application of an electric field to CQWs gives a good opportunity
for change of the spin relaxation anisotropy.

Our measurements are performed at photoexcited carrier concentrations about $(3 \ldots 5) \cdot 10^{10}$~cm$^{-2}$. For such density the percolation threshold is overcomed, so the maximum 
of the photoluminescence line corresponds to radiative recombination of free excitons.~\cite{Larionov2000} 
%the Kerr signal decays due to spin dephasing of free excitons.
% i.e. due to spin relaxation of electrons and holes bound in excitons. 
The electron-hole exchange interaction is suppressed due to spatial
electron and hole separation in CQWs as well as by a magnetic field. Since the hole spin relaxation is rather fast (some picoseconds) we conclude that the observed Kerr signal decay is caused by 
%separate spin relaxation of electrons and holes 
spin relaxation of electrons bound in excitons. 
This suggestion is correlated with the fact that pronounced Kerr rotation signal appears at voltage
$U> 0.2$~V. It could be explained if originally CQWs are
positively charged and one needs to apply some millivolts to
compensate an excess hole concentration. So the Kerr signal is caused by magnetization of the electrons orbitally coupled with holes while their spins are independent.

%We attribute our experimental data (Fig.~1) to macroscopic
%magnetization created only by electrons because of spatial
%electron and hole separation in CQWs and rather short (some picoseconds)
%hole spin dephasing time.  
%

{\bf Theory.}
%In the experiment, interwell excitons are resonantly excited. 
%Free motion of exciton means free motion of its electron and hole too. 
Under excitation of a free exciton its electron propagates in the structure plane with an average momentum ${\bm k}_e = {\bm K}m_e/M$, where ${\bm K}$ and $M$ are the momentum and effective mass of the exciton as a whole, and $m_e$ is the electron effective mass. Therefore SIA and BIA effective magnetic fields appear acting on the spin of an electron bound in exciton.
Due to exciton scattering randomly changing a direction of its wavevector ${\bm K}$,
the D'yakonov-Perel'-like spin relaxation takes place where electron spin precession is accompanied by exciton motional narrowing.~\cite{Silva&LaRocca97} As a result, spin relaxation time anisotropy is present due to interference of SIA and BIA.

For the D'yakonov-Perel' spin relaxation
mechanism the following expressions exist for the anisotropic
spin relaxation rates:~\cite{Averkiev02,Averkiev99}
\begin{equation}
\label{tau_s_DP}
	{1 \over \tau_z} = C(\alpha^2+\beta^2) \, \tau_p, \quad  {1 \over \tau_{x,y}} = {C\over 2} (\alpha \pm \beta)^2 \, \tau_p.
\end{equation}
Here the times $\tau_{x,y,z}$ are the relaxation times of the spin oriented along $x \parallel [1\bar{1}0]$,
$y\parallel [110]$, and  $z \parallel [001]$,  $\alpha$ and $\beta$ are Rashba and Dresselhaus electron constants, respectively, which determine the spin-orbit splittings of free electrons with the wavevector $\bm k_e$:
$\Delta_{\rm SIA} = 2\alpha k_e$, $\Delta_{\rm BIA} = 2\beta k_e$,
$\tau_p$ is the exciton momentum relaxation time, and $C = (m_e/M)^2 \overline{K^2}/\hbar^2$, where $\overline{K^2}$ is the mean value of the squared exciton in-plane wavevector.

Electron spin dynamics in the presence of an external magnetic field and for anisotropic spin
relaxation is described by the kinetic equation
\begin{equation}
    \label{S}
    {\partial {\bm S} \over \partial t} + {\bm S} \times {\bm \Omega} + \hat{\bm \Gamma} {\bm S} + {{\bm S} \over \tau_0}= 0.
\end{equation}
Here ${\bm S}$ is the electron spin density, ${\bm \Omega}$ is the Larmor frequency vector, and $\hat{\bm \Gamma}$ is the tensor of spin relaxation rates diagonal in the axes $x$, $y$ and $z$:
%$x \parallel [1\bar{1}0]$, $y\parallel [110]$, and  $z \parallel [001]$:
%
%The studied (001) grown CQWs have $C_{2v}$ symmetry which means that the tensor $\hat{\bm \Gamma}$ has
%three eigenvalues being the relaxation times of the spin oriented along $x \parallel [1\bar{1}0]$,
%$y\parallel [110]$, and  $z \parallel [001]$:~\cite{Averkiev99,Averkiev02}
$\Gamma_{xx} = {1/ \tau_x}$, $\Gamma_{yy} = {1/ \tau_y}$, and $\Gamma_{zz} = {1 / \tau_z}$.
%\[
%\Gamma_{xx} = {1\over \tau_x}, \quad \Gamma_{yy} = {1\over \tau_y}, \quad \Gamma_{zz} = {1\over \tau_z}.
%\]

If the electron spin is initially oriented along the axis $z$, the temporal behavior of the spin density
$z$-component is sought in the form $S_z(t) = S_z(0) \exp{(-t/\tau_0 + \lambda t)}$, where $\lambda$ is the complex
frequency.
Spin dynamics equation~\eqref{S} yields a cubic equation for $\lambda$:
\begin{eqnarray}
    \label{cubic_eq}
      &&  \lambda^3 + \left( {1 \over \tau_x} +{1 \over \tau_y} + {1 \over \tau_z} \right) \lambda^2 + \\
      &&  \left( {1 \over \tau_x \tau_y} + {1 \over \tau_y \tau_z} + {1 \over \tau_z \tau_x}
        \right) \lambda + {\Omega_x^2 \over \tau_x} + {\Omega_y^2 \over \tau_y} + {1 \over \tau_x \tau_y \tau_z}
        =0. \nonumber
\end{eqnarray}
Here $\Omega_{x,y}$ Larmor frequency projections on the axes $x$ and $y$, and  $\Omega_{z}=0$ because we consider a magnetic field lying in the structure plane.

Equation~\eqref{cubic_eq} has three roots, $\lambda_{1,2,3}$. One of them, $\lambda_1$, is real and does not describe
damping of spin oscillations. The imaginary parts of two others ($\lambda_2 = \lambda_3^*$) are equal by the
absolute value to the spin dephasing rate. This results in the following spin dynamics law:
\begin{equation}
\label{Sz}
    S_z(t) = A \, \exp{[-t(1/\tau_0 + 1/\tau_s)]} \, \cos(\Omega t + \varphi) ,
\end{equation}
which yields Eq.~\eqref{I} for time-resolved Kerr rotation  experiments.

If a magnetic field is strong enough: $\Omega \tau_{x,y,z} \gg 1$ or spin relaxation anisotropy is small:
$|1/\tau_x - 1/\tau_y| \ll 1/\tau_{x,y,z}, \Omega$, then one can derive an analytical expression
for the spin dephasing time $\tau_s$. In both limits we have the following result:
\begin{equation}
\label{tau_s}
    {1 \over \tau_s} = {1 \over 2} \left( {1 \over \tau_z} +{\sin^2{\theta} \over \tau_x} + {\cos^2{\theta}
    \over \tau_y} \right),
\end{equation}
where $\theta$ is the angle between magnetic field $\bm B$ and the axis $x \parallel [1 \bar{1} 0]$. {Here we ignore the effect of small $g$-factor anisotropy on the value of $\tau_s$ as well as small renormalization of the QB frequency.}

{\bf Discussion.}
Figure~1(b) demonstrates a fast growth of the spin dephasing time with increasing of applied bias  in  the range $U=0.5 \ldots 1.3$~V. This behavior is explained by the D'yakonov-Perel' spin relaxation mechanism which is dominant for electrons in GaAs-based semiconductor heterostructures. Indeed, scattering from interface microroughness is 
accelerated with decrease of the confinement size. The momentum relaxation time $\tau_p$ became shorter which suppresses spin relaxation according to Eq.~\eqref{tau_s_DP}. 
%{??? In addition, the radiative recombination time becomes longer with increasing of bias due to stronger spatial separation of electrons and holes. Since the radiative exciton lifetime is comparable with the Kerr signal decay time, such increase of the former also results in longer decoherence.}
The order of magnitude of $\tau_s \sim 1$~ns corresponds to a reasonable spin splitting constant $\alpha \approx 10^{-7}$~meV$\cdot$cm at $K=5 \cdot 10^5$~cm$^{-1}$, $m_e/M =0.3$, and $\tau_p = 1$~ps.

In our experiments the QBs are well pronounced, see Fig.~1(a). Therefore the relation $\Omega \tau_{x,y,z} \gg 1$ is true, and we can use Eq.~\eqref{tau_s} for description of the spin dephasing time angular dependence. It yields
\begin{equation}
\label{tau_s_fit}
    \tau_s(\theta) = {D \over 1 + b \cos{2\theta}},
\end{equation}
where $b < 1/3$.
The result of the experimental data fitting according to Eq.~\eqref{tau_s_fit} with  $D$ and $b$ as adjustable parameters 
is shown in Fig.~2 by solid lines. One can see a good agreement
between experiment and theory for all values of the bias. 

The performed quantitative description of the spin relaxation
anisotropy allows one to derive the ratio of SIA and BIA
electron spin-orbit splittings. 
The  parameter $b$ in Eq.~\eqref{tau_s_fit} determines the ratio of the Rashba  and
Dresselhaus constants:
\begin{equation}
\label{ratio}
    \left( {\alpha \over \beta} \right)^{\pm 1}= {3b \over 1 + \sqrt{1-(3b)^2}}.
\end{equation}
This expression demonstrates that Kerr rotation measurements do not allow one to distinguish between the constants $\alpha$ and $\beta$. Indeed, the symmetric form of Eq.~\eqref{tau_s_DP} yields  Eq.~\eqref{ratio}  either for the ratio $\alpha/\beta$ or for $\beta/\alpha$. However disappearance of the anisotropy at $U \approx 1.3$~V  means that one of the spin splittings nullifies. Since the Dresselhaus splitting can not be tuned to zero by bias, we conclude that it is the Rashba one, and Eq.~\eqref{ratio} describes namely the value of  $\alpha/\beta$ at 0.7~V $< U < 1.6$~V. 
We used Eq.~\eqref{ratio} for derivation of the ratio $\alpha/\beta$ and plotted
the result in Fig.~3 (see also the values given in Fig.~2). 
The values $\alpha/\beta > 1$ are taken at $U < 0.7$~V because the ratio should be monotonous function of the bias.

One can see strong variations of this parameter by the electric field: at $U>0.6$~V the ratio $\alpha/\beta$ decreases monotonously changing its sign at $U \approx 1.3$~V. 
%(the polar plot $\tau_s(\theta)$ is a circle at this bias)
At this bias the polar plot $\tau_s(\theta)$ is a circle. 
At higher bias the anisotropy is present again, but the ellipse is rotated by 90 degrees relative to that for low voltage.
\begin{figure}[t]
\begin{center}
\includegraphics[width=0.55\linewidth]{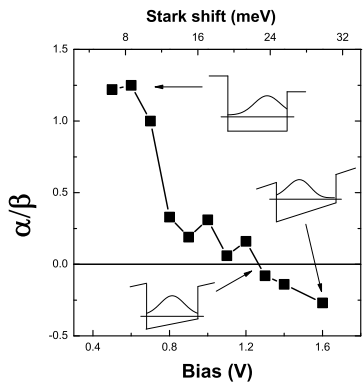}
\end{center}
\caption{The ratio of Rashba and Dresselhaus constants,
$\alpha/\beta$, as a function of an applied bias. Insets show the heteropotential profiles and the electron wavefunction in the lowest well of the CQW structure.}
\end{figure}
The observed behavior of the ratio $\alpha/\beta$ is mainly due to change of SIA by applied bias. The structure under study contains some built-in SIA at $U=0$ due to different properties of the interfaces which can be modeled by different height of the left and the right barriers (see insets to Fig.~3). External electric field suppresses this built-in asymmetry resulting in the decrease of $\alpha$. At $U \approx 1.3$~V the external and internal SIA compensate each other resulting in realization of the most symmetric confinement potential. This is  unambiguously indicated by the spin dephasing time isotropy at this voltage. At higher bias, the external contribution to SIA dominates which leads to change of the sign of the constant $\alpha$. The BIA constant $\beta$ is also sensitive to the external electric field because it is determined by the size of the electron confinement. However $\beta$ has the same sign at any bias therefore the ratio $\alpha/\beta$ changes mainly due to the dependence $\alpha(U)$ (so-called ``Rashba effect'').

The spin dephasing time saturation at high bias $U=1.3 \ldots 1.6$~V presented in Fig.~1(b) can be also explained by the above discussed compensation of SIA. At these voltages, increase of the momentum relaxation rate  competes with the suppression of the SIA spin splitting. As a result the spin dephasing time growth slows down in accordance with Eq.~\eqref{tau_s_DP}.

%\section{Conclusions}

To summarize, the electron spin dephasing time anisotropy is observed
in the biased (001) grown n-i-n CQWs. Spin decoherence is
studied by means of time-resolved Kerr rotation effect at
magnetic field differently oriented in the structure plane. The anisotropy is caused by interference of SIA and BIA spin-splittings in the electron spin relaxation via the D'yakonov-Perel' mechanism. It is demonstrated
that SIA is changed by bias according to the
Rashba effect in the studied structure. It is shown that spin-orbit
splitting can be controlled by electrical means in
n-i-n GaAs/AlGaAs coupled quantum wells.

%\section{Acknowledgements}

The authors thank M.\,M.~Glazov for fruitful discussions, and J.~Hvam and K.~Soerensen for preparation of the samples. This research is supported
by RFBR,  ``Dynasty'' Foundation --- ICFPM, and President grant for young russian scientists.

\end{document}